# Near-real-time global gridded daily $CO_2$ emissions 2021

**Authors**

Xinyu Dou[1], Jinpyo Hong[2], Philippe Ciais[3], Frédéric Chevallier[3], Feifan Yan[4], Ying Yu[5], Yifan Hu[6], Da Huo[1], Yun Sun[7], Yilong Wang[8], Steven J. Davis[9], Monica Crippa[10], Greet Janssens-Maenhout[10], Diego Guizzardi[10], Efisio Solazzo[10], Xiaojuan Lin[1], Xuanren Song[1], Biqing Zhu[1], Duo Cui[1], Piyu Ke[1], Hengqi Wang[1], Wenwen Zhou[11], Xia Huang[11], Zhu Deng[11], and Zhu Liu[1]*

**Affiliations**

1. Department of Earth System Science, Tsinghua University, Beijing, 100084, China

2. Department of Computer Science and Technology, Tsinghua University, Beijing, 100084, China

3. Laboratoire des Sciences du Climat et de l'Environnement, LSCE/IPSL, CEA-CNRS-UVSQ, Université Paris-Saclay, Gif-sur-Yvette, France

4. Key Laboratory of Marine Environment and Ecology, and Frontiers Science Center for Deep Ocean Multispheres and Earth System, Ministry of Education, Ocean University of China, Qingdao, 266100, China

5. Environmental Sciences and Engineering, University of North Carolina at Chapel Hill, Chapel Hill, USA

6. Key Laboratory of Sustainable Forest Ecosystem Management, Northeast Forestry University, Harbin, 150040, China

7. School of Environmental Science and Engineering, Tianjin University, Tianjin, 300072, China

8. Key Laboratory of Land Surface Pattern and Simulation, Institute of Geographical Sciences and Natural Resources Research, Chinese Academy of Sciences, Beijing, 100101, China

9. Department of Earth System Science, University of California, Irvine, CA, USA

10. European Commission, Joint Research Centre (JRC), Ispra, Italy

11. Product and Solution & Website Business Unit, Alibaba Cloud, Hangzhou, 311121, China

corresponding author(s): Zhu Liu (zhuliu@tsinghua.edu.cn)

**Abstract**

We present a near-real-time global gridded daily $CO_2$ emissions dataset (GRACED) throughout 2021. GRACED provides gridded $CO_2$ emissions at a 0.1° × 0.1° spatial resolution and 1-day temporal resolution from cement production and fossil fuel combustion over seven sectors, including industry, power, residential consumption, ground transportation, international aviation, domestic aviation, and international shipping. GRACED is prepared from a near-real-time daily national $CO_2$ emissions estimates (Carbon Monitor), multi-source spatial activity data emissions and satellite $NO_2$ data for time variations of those spatial activity data. GRACED provides the most timely overview of emissions distribution changes, which enables more accurate and timely identification of when and where fossil $CO_2$ emissions have rebounded and decreased. Uncertainty analysis of GRACED gives a grid-level two-sigma uncertainty of value of ±19.9% in 2021, indicating the reliability of GRACED was not sacrificed for the sake of higher spatiotemporal resolution that GRACED provides. Continuing to update GRACED in a timely manner could help policymakers monitor energy and climate policies' effectiveness and make adjustments quickly.

**Background & Summary**

Global climate change mitigation plans and efforts require countries, regions, cities and companies worldwide to set regional and local carbon emission control targets and emission reduction plans[1,2] and to monitor progress towards these targets over time. As the climate crisis becomes increasingly severe, countries and regions are expected to raise their climate ambitions and set a schedule for "carbon neutrality". As a climate mitigation management measure, "carbon neutrality" can effectively mitigate the global greenhouse effect. Carbon emission data is not only a critical tool for monitoring the progress toward carbon neutrality, but also an essential basis for assessing national carbon peaking levels, addressing climate change, and formulating corresponding climate policies. Timely, fine-grained gridded carbon emission data sets are particularly important for global climate change research[3,4]. Often, fine grained data are difficult to visualise over the globe, and clear visualisation tools are also needed.

Current global gridded $CO_2$ emissions datasets contain the Open Source Data Inventory for Anthropogenic $CO_2$ (ODIAC), the Community Emissions Data System (CEDS), and the Emission Database for Global Atmospheric Research (EDGAR)[5-8]. However, most of the datasets mentioned above have a data lag of at least one year, and do not present sub-monthly temporal changes associated with weather, climate seasonality, economic activity, and market shocks on energy production and trade.

In a previous study, we described GRACED, a near-real-time Global Gridded Daily $CO_2$ Emissions Dataset, for 2019 and 2020 for the first time, which can be updated at a 0.1° × 0.1° spatial resolution and 1-day temporal resolution[9]. In this study, the framework diagram is shown in Fig. 1. We present GRACED for 2021, including gridded daily fossil fuel $CO_2$ emissions on the basis of multi-source datasets: the near-real-time global daily national dataset of sectoral $CO_2$ emission due to fossil fuel consumption and cement production (Carbon Monitor), Emissions Database for Global Atmospheric Research (EDGAR) and spatiotemporal patterns of nitrogen dioxide ($NO_2$) retrieved from satellites. GRACED provides gridded $CO_2$ emissions in the following seven sectors: industry sector (incl. cement process), power sector, residential consumption sector, ground transport sector, international shipping sector, international aviation sector, and domestic aviation sector. By comparing with other global gridded $CO_2$ emission datasets, we further analyze spatiotemporal and sectoral uncertainties in $CO_2$ emissions over the period 2019-2021 to test the reliability of the estimated results.

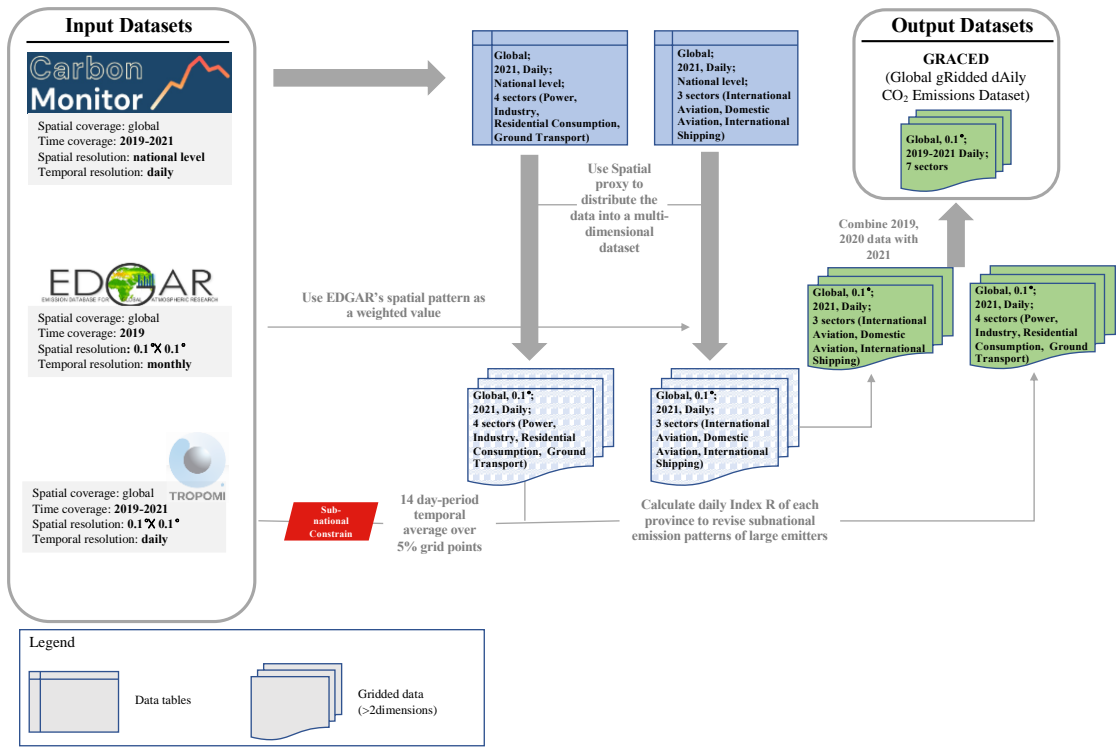

Fig. 1 Framework diagram of a top-down spatially gridding approach.

This research presents the latest analysis of near-real-time high-resolution gridded fossil $CO_2$ emissions for the year 2021. One of the advantages of our dataset is that it provides worldwide near-real-time monitoring of $CO_2$ emissions with different fine spatial scales at the sub-national level, such as cities, thus enhancing our comprehension of spatial and temporal changes in $CO_2$ emissions and anthropogenic activities. With the extension of GRACED time series, we present crucial daily-level input to analyze $CO_2$ emission changes during the COVID-19 pandemic, which will ultimately facilitate and aid in designing more localized and adaptive management policies for the purpose of climate change mitigation in the post-COVID era.

Our analysis of spatial distribution of emissions is a downscaling of Carbon Monitor daily national totals using EDGAR spatial distribution of activity data and TROPOMI NO2 measurements to infer changes in this spatial distribution. The gridded uncertainty of the GRACED dataset is also quantified. GRACED exhibited a grid-level two-sigma uncertainty value of ±23.1%, ±19.9%, and ±19.9% corresponding to 2019, 2020, and 2021. The primary source of GRACED uncertainty is from Carbon Monitor data. Also, comparison with other gridded $CO_2$ emission dataset revealed a relatively low uncertainty of GRACED dataset.

## Methods

**Input datasets.** *Carbon Monitor national-level emissions inventory.* Estimated GRACED emissions are based on the near-real-time daily $CO_2$ emissions from fossil fuel combustion and cement production provided by Carbon Monitor since January 1, 2019 (data accessible at https://carbonmonitor.org/)[3,4,10]. Carbon Monitor provides emissions at a national and sectoral level, drawing on near-real-time activity data and inventories for the reference year of 2019[3].

Since January 1, 2019, Carbon Monitor calculated daily national $CO_2$ emissions in five sectors (electricity, industrial production, ground transportation, home consumption, and domestic aviation) as well as daily international aviation and shipping emissions. These figures are provided for the nations, groups of countries, or regions listed below: China, the United States (US), India, the United Kingdom (UK), Italy, Germany, France, the rest of the European Union, Russia, Japan, Brazil, and the rest of the globe. Carbon Monitor uses near-real-time activity data, including hourly electricity generation data from 31 countries, traffic congestion data in 416 cities worldwide, daily sea and air transport activity data, monthly production data of steel, cement, and other energy-intensive industrial products for 62 countries or regions, and fuel use emission data of the previous year corrected based on daily air temperature fluctuation emitted by commercial and residential buildings. Overall, Carbon Monitor's input activity data covers about more than 70% of global industry and power emissions, 85% of ground transport emissions, and 100% of residential consumption and international bunker emissions, respectively. Carbon Monitor also provides the emissions as an aggregate for the rest of the world where data for some sectors (mainly electricity generation) is not directly available. National and worldwide daily $CO_2$ emissions with detailed information in six sectors and major nations were eventually computed by using these high temporal resolution data.

*EDGAR v5.0_FT2019 data.* In GRACED, emissions are spatially allocated using EDGARv5.0_FT2019 gridded spatial activity data underlying $CO_2$ emissions defined for sub-sectors in Intergovernmental Panel on Climate Change (IPCC) and novel geospatial proxies[7,8].

The European Union Emissions Database for Global Atmospheric Research (EDGAR) is an open-source platform developed and maintained by the European Environment Agency (EEA) and is widely used as the default for emission estimates in inventory systems. EDGAR supports the monitoring of the climate policy implementations, and in particular the Paris Agreement and significantly contributes to the quantification of IPCC national inventory guidelines and to the evaluation of the GHG budgets at various levels (http://verify.lsce.ipsl.fr/)[8]. EDGAR emissions data are split into sub-sectors specified by IPCC methodology and geospatial data / spatial proxy data like point source and line source location with the resolution of $0.1° \times 0.1°$ [7,8]. Driven by the rapid developments in scientific knowledge on the generation process of GHG emissions as well as the accessibility of more recent information, the EDGARv5.0_FT2019 dataset now includes new spatial proxies to allocate the emissions related to population based upon the Global Human Settlements Layer (GHSL) product [8,11].

We used the version EDGARv5.0_FT2019, which includes new spatial proxies to allocate population-related emissions based upon the Global Human Settlements Layer product[8].

EDGAR v5.0_FT2019 includes spatial activity data for all fossil $CO_2$ sources, including fossil fuel combustion, metal (ferrous and non-ferrous) production processes, non-metallic mineral processes (such as cement production), agricultural liming and solvents use, urea production [8]. Data are presented for every country, plus bunker fuels, at monthly intervals provided each major source category, and spatially distributed at a $0.1° \times 0.1°$ grid over the globe from 1970 to 2018 (https://edgar.jrc.ec.europa.eu/overview.php?v=50_GHG)[7,8].

*TROPOMI $NO_2$ retrievals data.* EDGAR employs static subnational patterns and currently do not extend after 2019. Static subnational patterns cannot reflect temporal changes in the spatial distribution of the emissions, for instance related to regional weather anomalies. COVID-19 has exacerbated this limitation for the year 2020 and 2021 as some regions experienced stronger confinements and emissions reductions than others even in the same country. Therefore, changes in the atmospheric distribution of a short-lived pollutant ($NO_2$), measured from satellite, co-emitted with fossil CO2 by combustion processes are used to reflect the changes in the human activities that cause $CO_2$ emissions[12].

We use the density $NO_2$ tropospheric vertical column retrieved from TROPOMI Level-2 products, including versions 1 and 2. Up until 5 August 2019, TROPOMI $NO_2$ retrievals had a ground resolution of $7 \times 3.5$ km$^2$, and afterwards at $5.5 \times 3.5$ km$^2$. Most of the cloudless locations of the globe are observed every day. As in our previous study, we perform aggregation in the form of daily $0.1° \times 0.1°$ maps using a quality assurance flag higher than 0.75 from the official retrieval[13] and average them over rolling 14-day periods. This procedure was performed to reduce the retrieval noises and reduce gaps[14].

**Method overview.** *Transforming the EDGAR categories into Carbon Monitor sectors.* We begin by mapping the Carbon Monitor emission categories to the EDGAR sectors based on the correspondence described in Table 1. We depend on EDGAR as the monthly spatial patterns to the fullest extent.

| No. | Carbon Monitor categories | EDGAR categories |
|---|---|---|
| 1 | Power | Oil_Power_Plants |
|   |       | Coal_Power_Plants |
|   |       | Gas_Power_Plants |
| 2 | Industry | Oil_combustion_and_use_for_Manufacturing |
|   |          | Gas_combustion_and_use_for_Manufacturing |
|   |          | Coal_combustion_and_use_for_Manufacturing |
|   |          | Cement_and_other |
| 3 | Residential consumption | Fossil_fuel_combustion_for_Human_Settlements |
| 4 | Ground transport | Fossil_fuel_combustion_for_Ground_Transport |
| 5 | International aviation | AIR Bunker oil for international transport |
| 6 | International shipping | SEA Bunker oil for international transport |
| 7 | Domestic aviation | AIR Bunker oil for domestic transport |

Table 1. Correspondence between Carbon Monitor categories and EDGAR categories used in this research.

In addition, in order to understand the sector classification adopted by GRACED in this study, the corresponding relationship between GRACED sector and IPCC sector is shown in Table 2.

| IPCC | IPCC description | This study |
|---|---|---|
| 1A1a | Public electricity and heat production | Power |

| 1A1bc | Other Energy Industries | Industry (incl. Cement Process) |
|---|---|---|
| 1A2 | Manufacturing Industries and Construction | Industry (incl. Cement Process) |
| 2A1 | Cement production | Industry (incl. Cement Process) |
| 1A3a | Domestic aviation | Domestic aviation |
| 1A3b | Road transportation no resuspension | Ground transport |
| 1A3c | Rail transportation | Ground transport |
| 1A3d | Inland navigation | Ground transport |
| 1A3e | Other transportation | Ground transport |
| 1A4 | Residential and other sectors | Residential consumption |
| 1A5 | Other Energy Industries | Residential consumption |
| 1C2 | Memo: International navigation | International shipping |
| 1C1 | Memo: International aviation | International aviation |

Table 2. The relationship between the 7 super-sectors of GRACED and the IPCC sectors.

*Spatial gridding process.* The second step entails a spatial gridding process. To downscale daily $CO_2$ emissions from Carbon Monitor, we utilize global monthly spatial patterns of $CO_2$ emission from EDGAR subsectors. The spatial distribution of $CO_2$ emissions was set to be unchanged compared to those of EDGAR in 2019. The nation and the time horizon for the adjustment will determine the validity of this assumption. Despite the rapid changes in subnational emissions within each country from 2019 to 2021, the impact of COVID-19 varied widely in time and magnitude across subnational areas. Consequently, for large countries being major emitters, we employ a sub-national proxy according to $NO_2$ tropospheric vertical column concentration retrieved from TROPOMI to assign national-level total $CO_2$ emissions into regional (first-level administration division) totals, followed by a 0.1° down-scaling according to the spatial patterns from EDGAR.

The GRACED spatial disaggregation system is illustrated in Fig. 1. It is a top-down approach that uses geographical patterns supplied by EDGAR, as well as a sub-national proxy according to TROPOMI $NO_2$ retrievals, in order to distribute Carbon Monitor daily emissions at country-level to our finer grid.

A comprehensive description of the methodology framework is elaborated as follows:

(1) First, use daily emissions at national level of the Carbon Monitor's seven sectors (Industry sector, Power sector, Residential Consumption sector, Ground Transport sector, International Aviation, Domestic Aviation, and International Shipping sector, see Table 1) to generate gridded daily emissions with monthly spatial patterns from EDGAR, additional correction can be performed under monthly spatial patterns. In particular, for the international aviation and international shipping sectors, daily emissions at the global level from Carbon Monitor are directly distributed with monthly spatial proxy from EDGAR.

$$GRACED\_v1_{c,d,s} = CM_{country,d,s} \times \frac{E\_EDGAR_{c,m,s}}{\sum_{i=1}^{n} E\_EDGAR_{i,m,s}} \quad (1)$$

Where $GRACED\_v1_{c,d,s}$ represents the first step / version value of estimated $CO_2$ emission for cell c, date d and sector s, $CM_{country,d,s}$ represents the Carbon Monitor's value of country *country* which grid cell c belongs to, for date d and sector s. n is the aggregate of number of grid cells within the country *country*. $E\_EDGAR_{c,m,s}$ represents the gridded $CO_2$ emission value from EDGAR for grid cell c, month m which date d belongs to and sector s.

(2) Subnational emission patterns for major emitters can change from year to year, with far-reaching implications for global emissions. This was apparent in 2021, with regional differences in changes induced by COVID-19, such as between the eastern and western United States and between southeastern and northwestern China. Capturing these emission changes at sub-national level is critical. However, these differences are not resolved by Eq. 1, which use static subnational emission patterns. According to reports, worldwide emission variations are broadly consistent with worldwide $NO_2$ column changes derived from satellite data[12] although the lifetime of $NO_2$ emissions varies with weather and locations. As a result, we assume that subnational emission changes follow the variations of satellite $NO_2$ concentration retrievals variations from 2019 to 2021. In particular, we compute an index R for each first-level administrative unit in major emitting countries. R is the mean $NO_2$ column concentration for each first-level administrative unit based on TROPOMI $NO_2$ column concentration data in year y.

$$R_{u,y} = NO_{2_{u,y}} \qquad (2)$$

Where u represents first-level administrative unit; y represents the year. $NO_{2_{u,y}}$ represents the temporal average of satellite $NO_2$ column concentration for the first administrative unit u in year y over a rolling 14-day period (as mentioned above) and spatially over the 5% of grid cells within each first-level administrative unit that have the highest annual $NO_2$ concentration average. The selection of the 5% largest mean values enables the filtering of patterns in emitting areas or in their immediate vicinity from the long range transport of $NO_2$. Next step, we delete negative spurious $NO_2$ concentration values for the 5% largest grid cells over the year 2019 and 2021 that may be generated and re-attribute the mass gain to the other 5% grid cells. Using TROPOMI $NO_2$ retrievals data, we then compute the index *R* for each first-level administrative unit in 2019 (2021).

Next, we generate $GRACED\_v2_{u,d,s1,2021}$, the daily emission value of a first-level administrative unit for sector *s1* in day d of 2021 modified by the TROPOMI $NO_2$ concentration in day d, which corresponds to the daily emission at national level from Carbon Monitor according to Eq. 3. *s1* includes the Industry sector, Power sector, Residential consumption sector, and Ground transport sector. *s2* incorporates the International shipping sector, International aviation sector, and Domestic aviation sector. For sector *s2* with more fixed spatial distribution patterns from 2019 to 2021, their emissions are not modified by TROPOMI $NO_2$ concentration to avoid causing spatial discontinuities.

$$GRACED\_v2_{u,d,s1,2021} = \frac{GRACED\_v1_{u,d,s1,2019} \times \frac{R_{u,2021}}{R_{u,2019}}}{\sum_{u=1}^{nu} GRACED\_v1_{u,d,s1,2019} \times \frac{R_{u,2021}}{R_{u,2019}}} \times CM_{country,d,s1,2021} \qquad (3)$$

Where $GRACED\_v1_{u,d,s1,2019}$ denotes the GRACED Version 1 emission estimate value for first-level administrative unit u in date d of 2019 and for sector *s1*. The total number of first-level administrative units in the country is denoted by *nu*. The first step is to compute $\frac{R_{u,2021}}{R_{u,2019}}$, the change ratio of R index in 2021 over 2019. We multiply $GRACED\_v1_{u,d,s1,2019}$, the first-level administrative unit emission estimate aggregated from GRACED Version 1 for year 2019, to update the first-level administrative unit emission value for 2021. The last step is to divide the updated first-level administrative unit emission estimate by the sum of the updated first-level administrative unit emission value $\sum_{u=1}^{nu} GRACED\_v1_{u,d,s1,2019} \times \frac{R_{u,2021}}{R_{u,2019}}$ in 2021 to perform the normalization in Eq. 3. As a result, the sum of emissions of the updated first-level administrative unit within a country matches the emission total at national-level from Carbon Monitor of 2021 after multiplying the emission value $CM_{country,d,s1,2021}$ at the national-level from Carbon Monitor.

Then, on the basis of updated first-level administrative unit emission $GRACED\_v2_{u,d,s1,2021}$ in 2021, we finalise the disaggregation of emissions with the spatial patterns from EDGAR data to distribute the emission estimate of each first-level administrative unit for major emitters to calculate the final version of gridded emission value $GRACED\_v2_{c,d,s1}$.

$$GRACED\_v2_{c,d,s_1} = GRACED\_v2_{u,d,s1,2021} \times \frac{E\_EDGAR_{c,m,s1}}{\sum_{i=1}^{n} E\_EDGAR_{i,m,s1}} \qquad (4)$$

Where *n* denotes the total number of cells belonging to this first-level administrative unit.

After revising and adjusting the gridded emissions of major emitters such as China, US, India, Japan, Brazil, UK, Germany, Spain, Italy, and France in 2021, GRACED of 2021 is finally produced.

## Data Records

**Spatial distribution of global $CO_2$ emissions.** Based on activity data of fossil fuel and cement production, GRACED provides fine-grained $CO_2$ emissions data at a $0.1° \times 0.1°$ spatial resolution and 1-day temporal resolution. Figure 2 reports the updated global gridded daily average $CO_2$ emissions in 2021, which identifies specific emission variations induced by sub-national emission allocation. According to Fig. 2, we can find that the spatial distribution of the average daily $CO_2$ emissions in 2021 is characterized by significant regional differences. The high values are concentrated in high-income or rapidly developing areas such as the south-eastern China, Western Europe, eastern U.S., India, Japan, and Korea. In contrast, low values are mainly found in eastern Russia, Africa, and central South America with low population density or low-income levels. In 2021, the emission grid in which the maximum value appears is 231741 tons of carbon per day per grid. The daily average total emissions are estimated to be about 4092 kg of carbon per day (kgC/d) per grid, which is higher than that of 2020 (3830 kgC/d per cell).

It is interesting to study the spatial distribution characteristics of $CO_2$ emissions over large cities. As shown in Fig. 2, the eight selected cities including Beijing, Shanghai, Tokyo, Los Angeles and the Greater London area are particularly visible on the daily $CO_2$ emission gridded map. By mapping the daily average $CO_2$ emissions of hot spots, we can find that the high values mainly coincide with densely populated urban areas or industrial areas. This is because sectors closely associated to human activities, such as industrial production and transportation, are usually energy-intensive and therefore contribute to a large amount of $CO_2$ emissions.

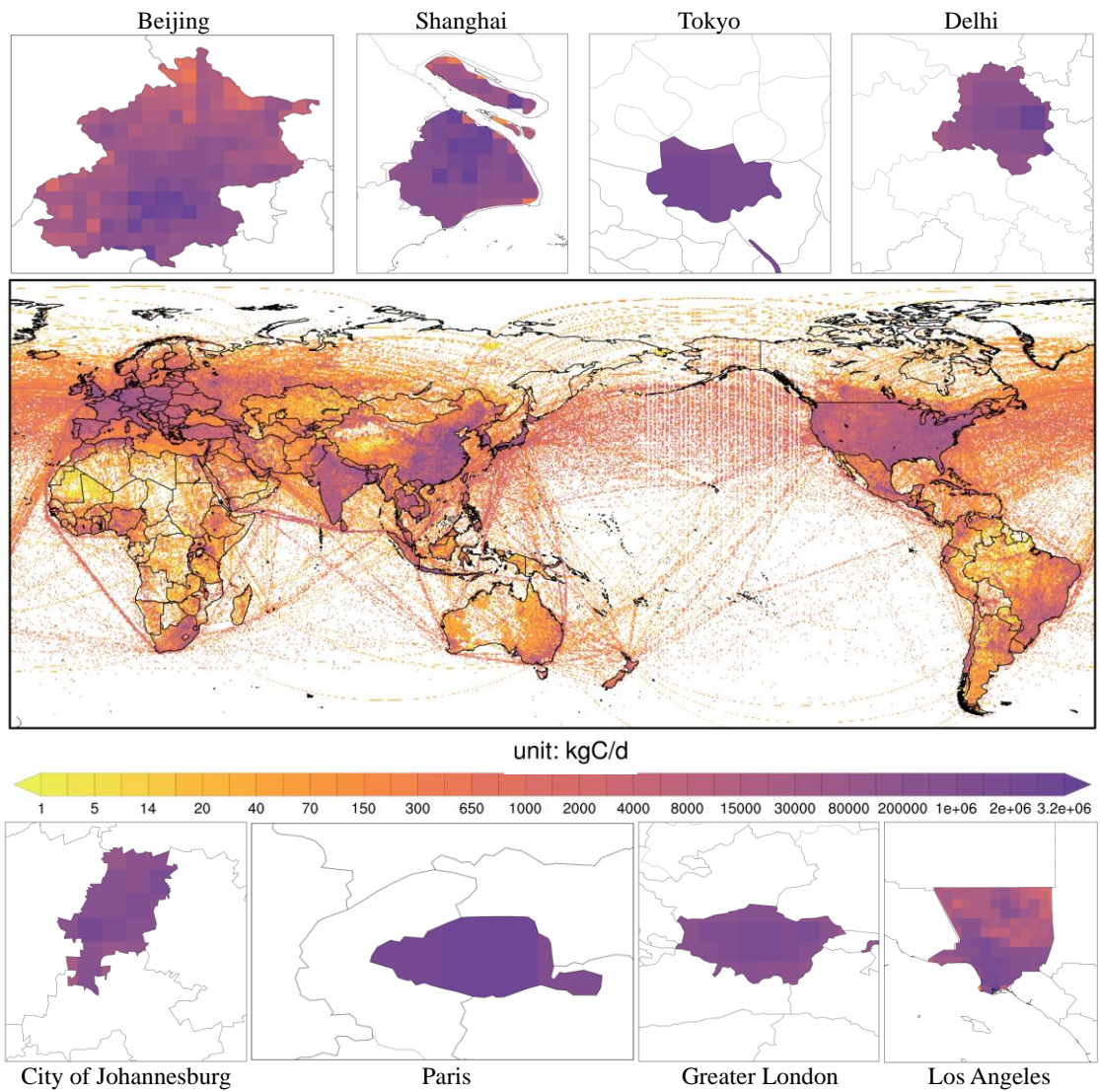

Fig. 2 The spatial $CO_2$ emissions distributions from GRACED in 2021 (unit: kgC/d per grid).

**Temporal patterns of global $CO_2$ emissions.** *Emission reductions between weekends and weekdays.* Figure 3 presents the temporal changes of the difference in emissions between weekends and weekdays in 2021. According to Fig. 3, we can find that in general, the global average weekend emissions are lower than the weekday emissions, with an average difference of -241 kgC/d per grid. The difference between weekday and weekend emissions is more pronounced in high-income regions such as Europe (dark blue areas) than in low-income regions such as Africa (light red or light blue areas). It is noteworthy that there are dark red areas in some states in the U.S. and parts of southern China, where the average weekend emissions in these regions are higher than the average weekday emissions. Linear regression shows that the spatial and temporal distribution characteristic is significantly correlated with emissions from the ground transportation sector, implying that reducing driving activities and recreations on weekends have a significant impact on weekend emissions reduction. In contrast, dark red areas are also found in coastal populated areas, which indicates that average weekend emissions in the international shipping and aviation sectors are generally higher than average weekday emissions.

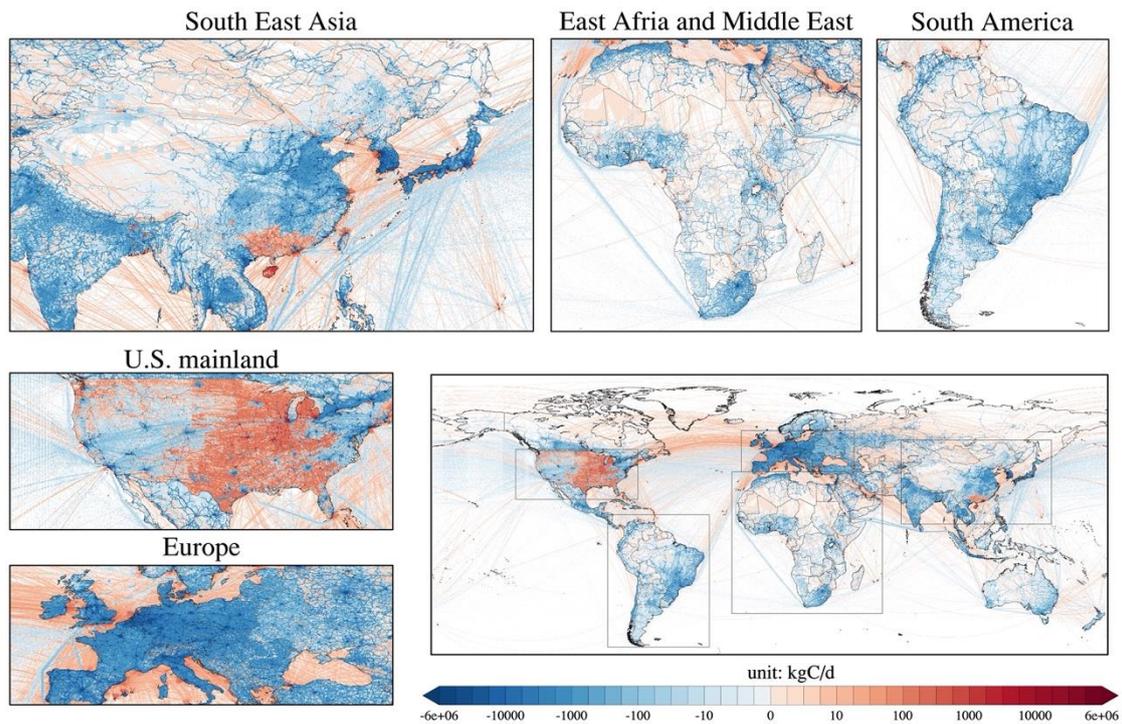

Fig. 3 Emission difference between weekends and weekdays map in 2021 (weekend minus weekday emissions).

For both 2020 and 2019, the global average weekday $CO_2$ emission was higher than weekend $CO_2$ emission generally[9]. The global average emission difference is -286 kgC per day per cell in 2020, which is larger than in 2021 (-241 kgC/d). Due to the economic recovery in 2021, general human travel has greatly increased on weekends compared to 2020, thus reducing the emission difference between weekdays and weekends.

*Difference between quarterly mean emissions.* Figure 4 reports the difference between the average daily emissions for each quarter[1] and the average daily emissions for the year in 2021. Figure 4 shows the four quarterly emission difference maps after removing the annual average emission value from the average emissions of each quarter. In this case, the first quarter is well above the annual average emission level, with a global average gap of 174 kgC per cell per day; and the second quarter is well below the annual average emission level, with a global average gap of -179 kgC per cell per day. This is mainly because energy consumption due to heating demand in the Northern Hemisphere versus cooling demand in the Southern Hemisphere makes $CO_2$ emissions from January to March higher than the rest of the year.

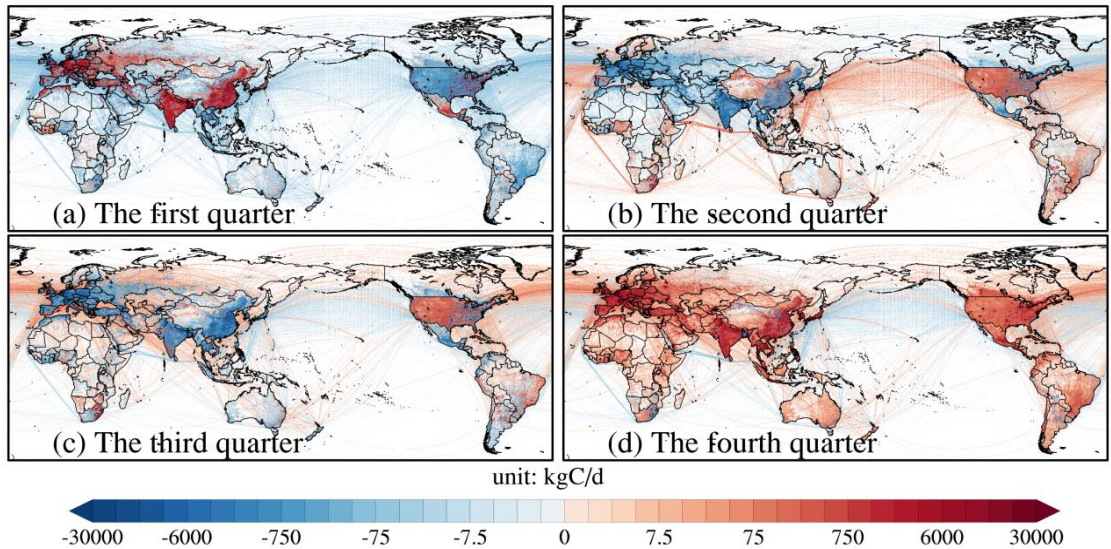

Fig. 4 Map of quarterly average minus yearly average emissions in 2021 (Note: remove the yearly average from each quarter).

**Emission variations between 2021 and 2020.** The maximum reduction date is defined as the date of the greatest emission decrease in 2021 compared with 2020, which reflects the day of the year when significant emission reduction was caused by the remaining impact of the new outbreak in 2021 (Fig. 5a). The green areas show where countries or regions around the world were mainly affected by the new wave of outbreaks in the early part of 2021. By contrast, the purple areas in China indicate economic stagnation caused by the large-scale lock-down policies in response to the Delta and Omicron variant outbreaks in late 2021.

Similarly, the maximum rebound date is defined as the date of the largest emission increase in 2021 compared with 2020, the day of the year experiencing the largest emission recovery in 2021 (Fig. 5b). The dark green area implies that China experienced the earliest rebound in 2021, while most countries or regions had the greatest rebound in the middle of 2021, reflected by the light green and purple areas. From a sectoral perspective, we find that the international shipping and international aviation sectors experienced the largest rebound in the second half of 2021, which indicates a later recovery in these two sectors.

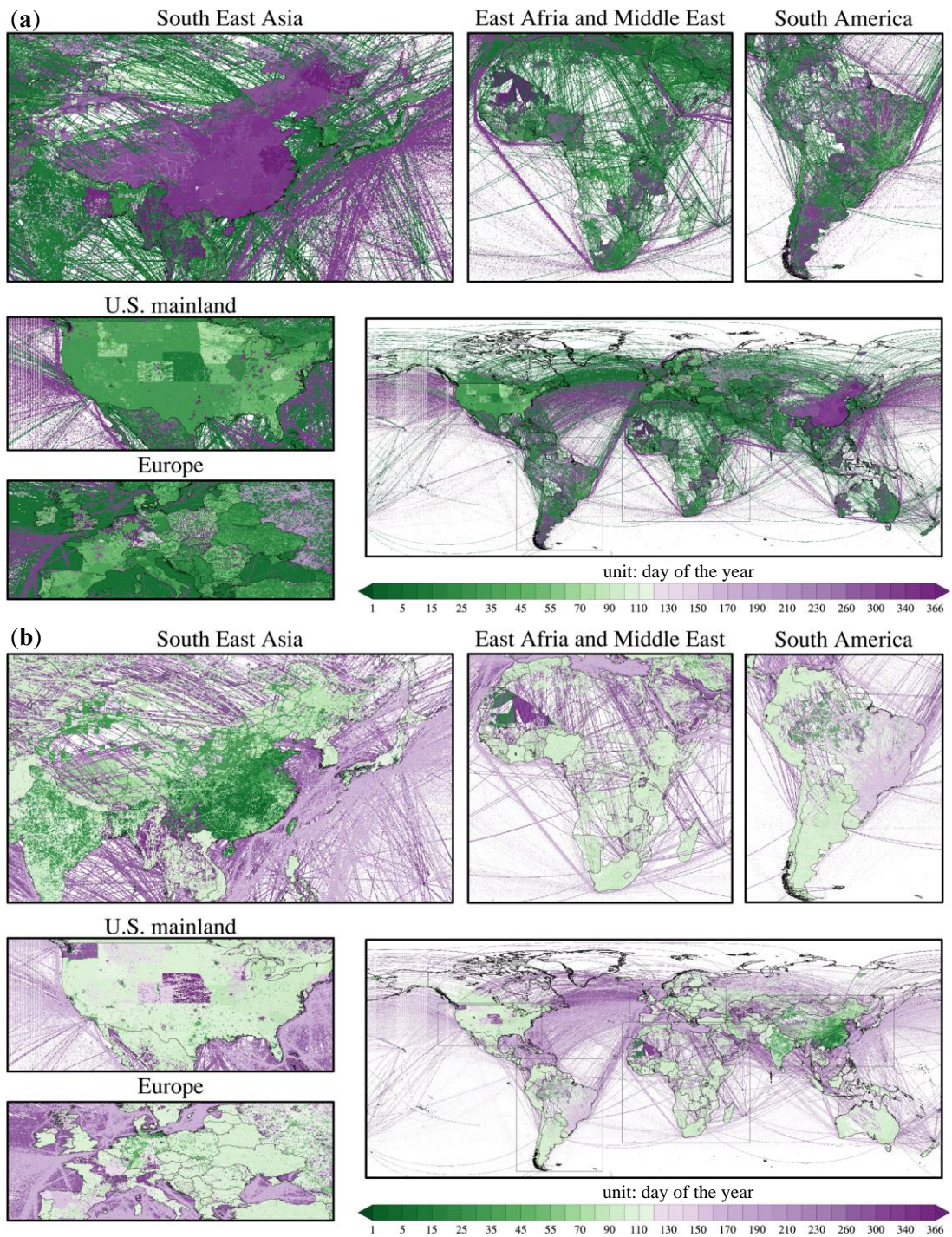

Fig. 5 The day of the year in 2021 with the maximum reduction (a), and the maximum rebound (b) of each grid in 2021 compared with 2020.

**Emission distribution analysis.** By analyzing the national distribution of emissions from a grid-wise perspective for US and China and plotting it in a natural logarithmic scaled histogram, GRACED provides insight into COVID-19's impact on national emissions of these two nations. The analysis results are consistent[10].

A decrease in the number of high emission grids (Ln value of 14 to 18) in the US for 2020 shows a notable impact of COVID-19. Among the six sectors, excluding the international shipping sector, mobility-related emissions have decreased more rapidly. International aviation, domestic aviation, and ground transport sectors display a severe plunge in the number of high emission grids in 2020, which aligns with the Carbon Monitor's result of 45.8%, 30.4%, and 9.6% decrease in the following sectors for 2020[15]. On the other hand, the decline in the residential consumption sector was relatively small, where the number of grids remained relatively unchanged. Pandemic-induced lockdown resulted in longer time spent home, negating the impact of COVID-19 on the residential consumption emission. The 2021 distribution indicates no difference or even increase in the number of high emission grids in every sector, excluding aviation-related sectors, reflecting the $CO_2$ emission rebound originating from the disappearance of lockdown, recovery of the economy, and increase of demand in power and industry sector.

China also exhibits a decrease in mobility-related emissions due to strict lockdown policies during the first half of 2020. However, due to the recovery of other sectors in the second half of 2020, China's distribution of the total emission has a different characteristic from that of the US. China's power and industry sector distribution shows an increase in the number of high emission grids, leading to a 0.9% increase in national $CO_2$ emissions[10]. In 2021, the disappearance of COVID-19 threats due to the zero-COVID policy led to a full recovery of emissions of every sector to the pre-pandemic level except the international aviation sector. The delayed recovery of the international aviation sector originates from strict quarantine policies.

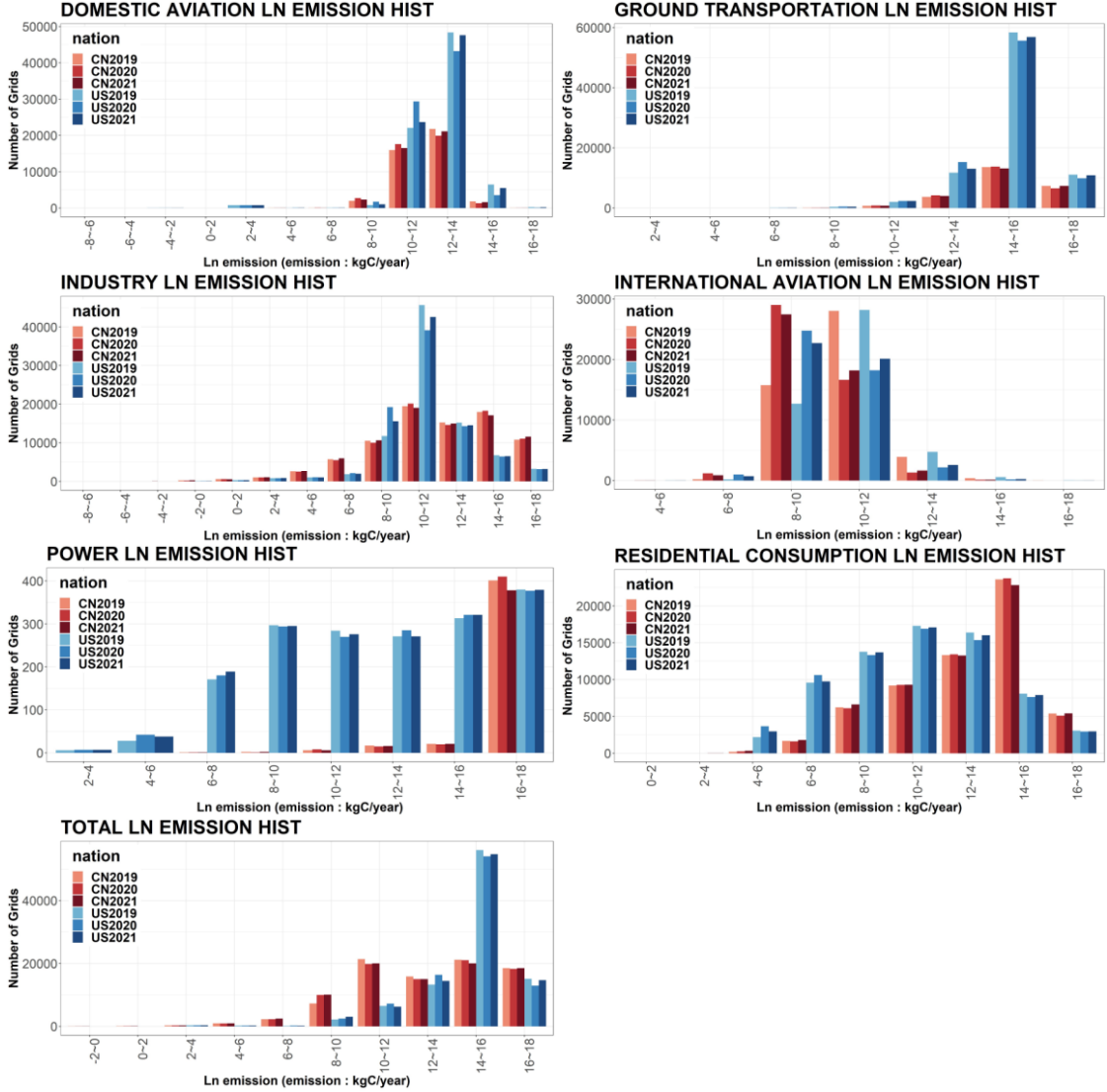

Fig. 6 Distribution of grid level emission for US and China from 2019 to 2021. The x-axis represents different bins of emissions values at grid cell level (Ln transformed because of the strong right-hand skewness of emissions distributions).

## Technical Validation

**Uncertainty analysis.** Uncertainty associated to each variable propagates to the final uncertainty of GRACED data. By applying the rule of error propagation and IPCC 2006[16,17] uncertainty analysis on Eq. 1 and Eq. 4, the following equations were obtained:

$$U_{GRACED\_v2_{c,y,s1}} = \sqrt{U_{CM_{u,y,s1}}{}^2 + U_{E\_EDGAR_{c,s1}}{}^2} \qquad (5)$$

$$U_{CM_{u,y,s1}} = \sqrt{U_{CM_{country,y,s1}}{}^2 + U_{R_{u,y}}{}^2} \qquad (6)$$

$$U_{GRACED\_v2_{c,y,s1}} = \sqrt{U_{CM_{country,y,s1}}^2 + U_{R_{u,y}}^2 + U_{E\_EDGAR_{c,s1}}^2} \qquad (7)$$

$$U_{GRACED\_v1_{c,y,s2}} = \sqrt{U_{CM_{country,y,s2}}^2 + U_{E\_EDGAR_{c,s2}}^2} \qquad (8)$$

$$U_{total} = \frac{\sqrt{\sum_{i=0}^{6}(U_{c,y,sr_i} \times X_{c,y,sr_i})^2}}{\sum_{i=0}^{6} X_{c,y,sr_i}} \qquad (9)$$

Where $U_x$ format indicates the uncertainty of variable $x$, $sr_0 \sim sr_6$ indicates the seven individual sectors that consist the total emission, and $X_{c,y,sr_i}$ is the annual sum of total emission for grid cell $c$, year $y$ and sector $sr$ in GRACED. Due to the nature of Carbon Monitor data, as mentioned[4] correlation between sectors were not considered in the uncertainty analysis of this research.

The methods utilized to calculate each uncertainty variable in the Eq. 7 and Eq. 8 are presented below:

For $U_{E\_EDGAR_{c,s}}$, we utilized the method proposed by Dai, et al. [18]. Probability density functions (PDF) of the values of $E\_EDGAR_{c,s}$ were fitted and the validity of the fitness were verified by Kolmogorov-Smirnov test. If the KS test showed significance in the fitted PDF, the fitted PDF was used to produce 100,000 simulated sample grids by Monte Carlo simulation. Subsequently, the cumulative distribution function (CDF) of the simulated grids and the original grids were compared and the difference between the two functions were used to calculate the uncertainty corresponding to the weighted grid value.

$$U_v = \frac{SCDF_v - TCDF_v}{SCDF_v} \times 100(\%) \qquad (10)$$

Where $U_v$ is the uncertainty corresponding to an emission amount $v$, $SCDF_v$ is the simulated CDF value for emission amount $v$ and $TCDF_v$ is the true CDF value for emission amount $v$.

Based on the equation and the original grid value, uncertainty value of each grid was calculated:

$$U_{E\_EDGAR_{c,s}} = \frac{SCDF_{E\_EDGAR_{c,s}} - TCDF_{E\_EDGAR_{c,s}}}{SCDF_{E\_EDGAR_{c,s}}} \times 100(\%) \qquad (11)$$

For $U_{CM_{country,y,s}}$, we assumed that the $U_{CM_{country,y,s}}$ will be equal to the uncertainty of the corresponding sector $s$ reported by previous studies of Carbon Monitor performed on the global scale regardless of the nation[3,10].

For $U_{R_{u,y}}$, as this research went through a pre-processing process for the TROPOMI NO$_2$ data to calculate $R_{u,y}$, we stated that the uncertainty of $R_{u,y}$ will be a constant value throughout the gridded dataset based on the result of Goldberg, et al.[19], and van Geffen, et al.[20]. Cooper, et al. [20] stated that the uncertainty of NO$_2$ are less than 5%, van Geffen, et al. [20] stated that the uncertainty originating from the slant column density (SCD) method in TROPOMI NO$_2$ is 10%, and Goldberg, et al. [19] mentioned that TROPOMI product have a low bias of 20% in urban areas.

$$U_{R_{u,y}} = \sqrt[3]{U_{Cooper} * U_{Geffen} * U_{Goldberg}} = \sqrt[3]{5 * 10 * 20} = 10(\%) \qquad (12)$$

Where $U_{Cooper}, U_{Geffen}, U_{Goldberg}$ denote the uncertainty mentioned in Cooper, et al. [21], van Geffen, et al. [20] and Goldberg, et al.[19].

The uncertainty of each sector was calculated based on the equations mentioned previously. As shown in Fig. 7, the smallest uncertainty in 2021 was obtained from the power sector (±13.5%), where the majority of this uncertainty originates from the uncertainty of Carbon Monitor's power sector data (±10%)[3]. The average value of uncertainty in grids for industry, residential consumption, and ground transport, other three sectors included in $s_1$, were ±33.4%, ±42.0%, and ±15.5%, respectively. On the other hand, average gridded uncertainty in international shipping sector, international aviation sector, and domestic aviation sector had a smaller uncertainty of ±16.7%, ±16.0%, and ±18.6%. The uncertainty values of EDGAR's spatial proxies calculated were all in the range of ±3.2% ~ ±11.7%. Relatively small uncertainty value of EDGAR's spatial proxies compared to GRACED's indicates that the main difference between GRACED's sectoral uncertainty originates from the difference in Carbon Monitor uncertainty for each sector. For instance, residential sector, the sector with the highest uncertainty of ±42.0% for the year 2020 and 2021, has more than 90% of its uncertainty originating from Carbon Monitor (±40%)[3], while contribution of EDGAR's residential uncertainty (±4.3%) is minimal. Also, there was a distinguishable difference in the average uncertainty between $s_1$ and $s_2$ sectors, which originates from higher value of Carbon Monitor uncertainty of $s_1$ sectors as well as R uncertainty included in the calculation procedure for $s_1$ sectors.

| GRACED | 2019 | 2020 | 2021 |
|---|---|---|---|
| Power | ±11.2% | ±13.4% | ±13.5% |
| Industry | ±18.4% | ±33.4% | ±33.4% |
| Residential | ±16.1% | ±42.0% | ±42.0% |
| Ground transport | ±14.1% | ±15.5% | ±15.5% |
| Domestic Aviation | ±16.1% | ±18.6% | ±18.6% |
| International Aviation | ±37.1% | ±16.0% | ±16.0% |
| International Shipping | ±16.7% | ±16.7% | ±16.7% |
| Total | ±23.1% | ±19.9% | ±19.9% |

Fig. 7 Annual sectoral uncertainty of GRACED.

Comparing the $CO_2$ emission uncertainty of 2019, 2020, and 2021, the uncertainty value is highest in 2019 (Fig. 7), while similar in 2020 and 2021. Higher average uncertainty in 2019 is induced by a larger Carbon Monitor uncertainty value for international aviation in 2019 (±34.2%) compared to 2021 (±10.2%)[3], resulting in the majority of grids located at sea having a significant uncertainty value difference from the other two years (Fig. 8). Closer observation indicates a similar grid distribution for 2020 and 2021 due to the spatial proxy distribution and constant Carbon Monitor uncertainty for 2020 and 2021. The areas with high uncertainty values are mainly distributed in Northwest Africa. These grids have small emission values for multiple sectors affecting the high uncertainty value of EDGAR for those corresponding sectors.

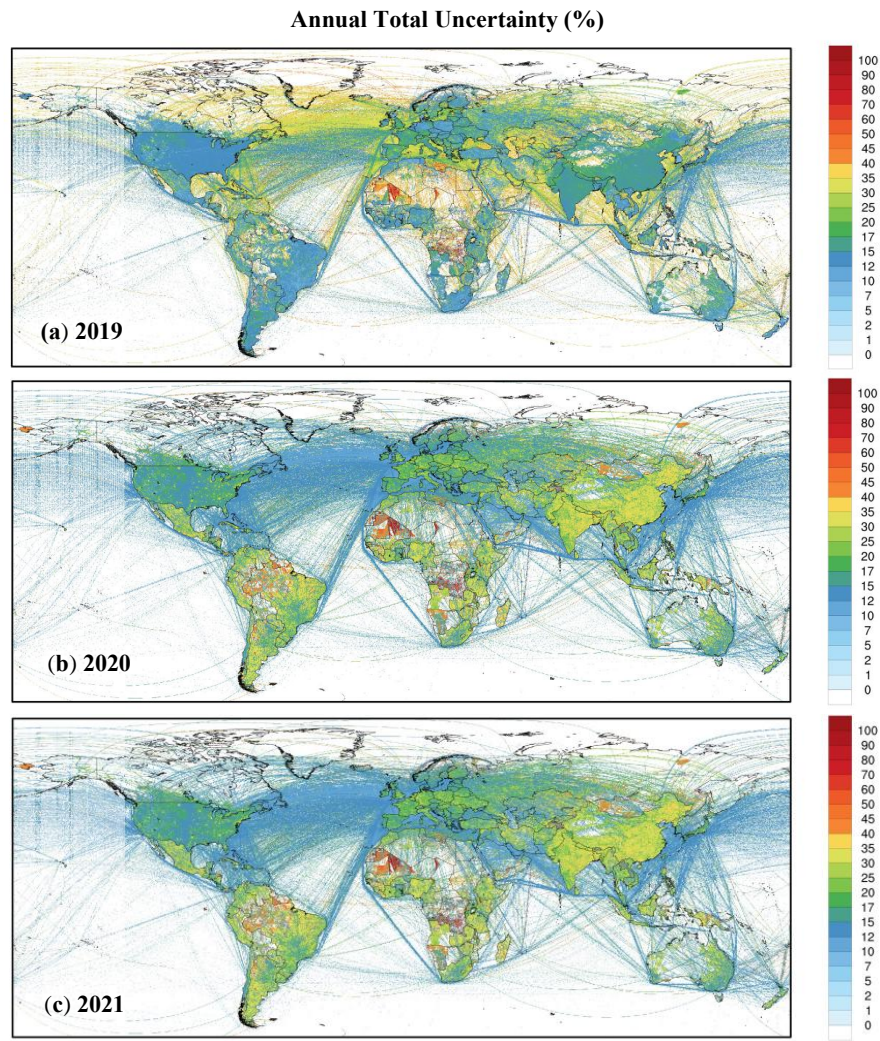

Fig. 8 Uncertainty of $CO_2$ emission of GRACED for every sector throughout 2019-2021.

The gridded sectoral uncertainty of 2021 displays additional information about the distribution of the uncertainties (Fig. 9). International aviation, power, and domestic aviation sector exhibited a relatively constant uncertainty in every grid due to even distribution in the corresponding sectors' EDGAR spatial proxy uncertainty. However, the ground transport and residential consumption sector displayed a relatively high uncertainty in parts of Africa. This issue originates from low emission values of EDGAR data grids, causing higher spatial proxy uncertainty.

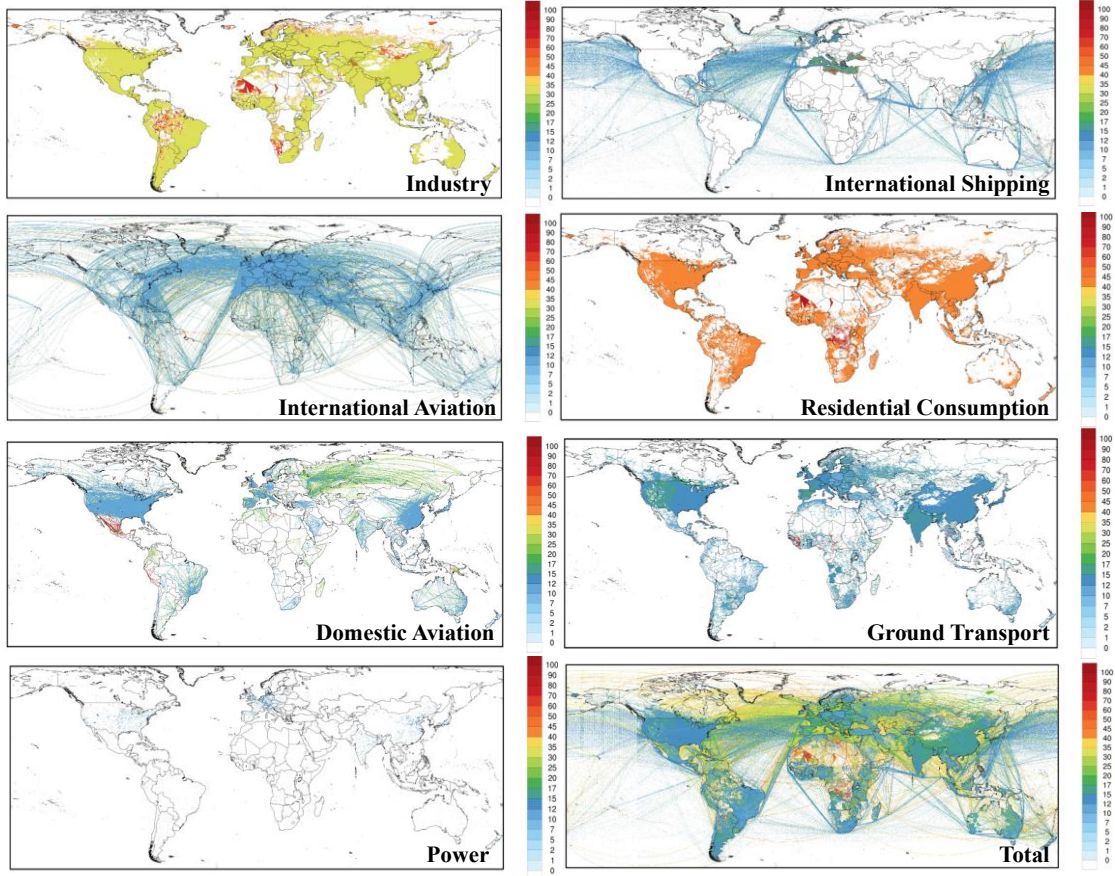

Fig. 9 Uncertainty of sectoral $CO_2$ emission in 2021.

**Validation against other datasets.** We also examined the distribution of emission in a grid-wise perspective for major emission datasets, GCP-GridFED[22], ODIAC[23] and EDGAR[24], and compared it with GRACED (Fig. 10). As GRACED employs extensive use of the EDGAR dataset, utilizing it as the main spatial proxy during the calculation procedure, the distribution of emission is very similar between GRACED and EDGAR throughout every emission range. This also originates from the total emission of Carbon Monitor's small difference from EDGAR's total emission. GRACED's distribution is displayed in accordance with GCP in the middle range (8-16) where the majority of the grids are included. The variance was only observed in the two extremes. The similarity in emission distribution and the number of non-zero emission grids were observed in GRACED, EDGAR and GCP-GRIDFED. However, ODIAC showed huge variance with the other three datasets in every range of emission. ODIAC showed a higher number of grids than the other datasets in every emission range. This huge variance could be originating from the calculation method of ODIAC which employs nighttime light data[25]. The nightlight-based approach of ODIAC can cause higher calculation values in the dataset compared to emission inventories for certain cities, due to a poor correlation between anthropogenic activities and nightlight intensity[26].

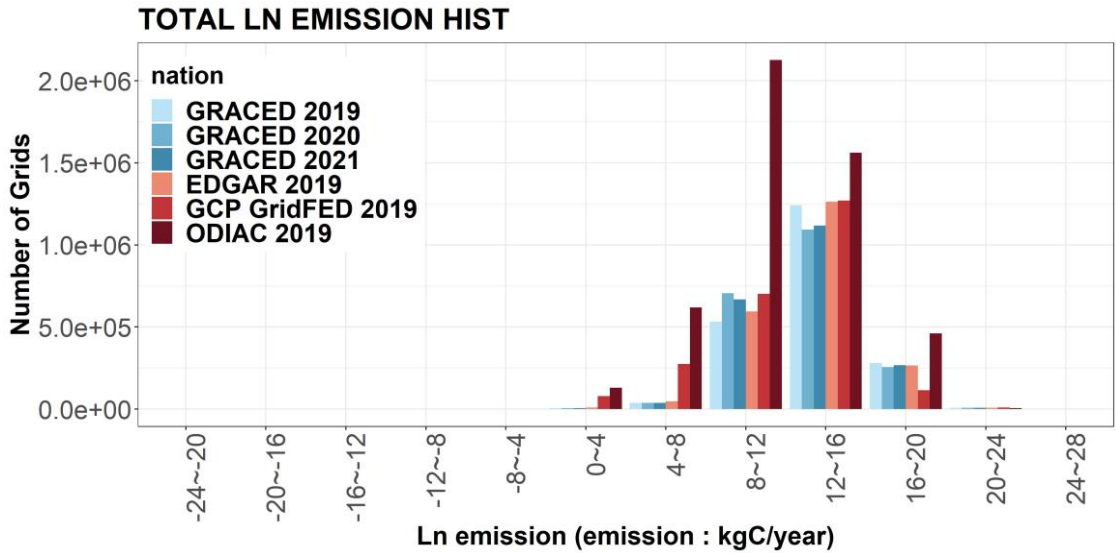

Fig. 10 Distribution of grids' emission for multiple emission datasets.

The chart below (Table 3) shows information about other published gridded $CO_2$ emission datasets. The majority of these datasets, including GRACED, have a spatial resolution of 0.1° × 0.1°. GRACED's spatial coverage also showed an equal level of other datasets. As mentioned previously, GRACED is the only dataset to provide near-real-time emission data in a temporal resolution of 1-day. Due to the relatively recent creation dataset of GRACED, the time coverage is shorter than other datasets. This issue will gradually be solved in the future as data accumulates.

| Dataset | GRACED | EDGAR v6.0[24] | ODIAC[23] | CEDS[11] | GCP-GridFED[27] |
|---|---|---|---|---|---|
| Spatial Resolution | 0.1° × 0.1° | 0.1° × 0.1° | 1° × 1° | 0.5° × 0.5° | 0.1° × 0.1° |
| Spatial Coverage | Global | Global | Global | Global | Global |
| Temporal Resolution | 1-day | 1-month (2000-2018) 1-year (1970-2018) | 1-month | 1-month | 1-month |
| Time Coverage | 2019-2022 | 1970-2018 | 2000~2019 | 1750-2019 | 1959-2019 |
| Classification | Industry, Power, Residential consumption, Ground transportation, Domestic aviation, International aviation, International shipping | Power Industry, Oil refineries and Transformation industry, Combustion for manufacturing, Aviation climbing & descent, Aviation cruise, Aviation landing & takeoff, Aviation supersonic, Road transportation no resuspension, Railways, pipelines, off-road transport, Shipping, Energy for buildings, Fuel exploitation, Non-metallic minerals production, Chemical processes, Iron and steel production, Non-ferrous metals production, Non-energy use of fuels, Solvents and products use, Agricultural waste burning, Agricultural soils, Solid waste incineration, Fossil Fuel Fires | Land, International bunker | Agriculture, Energy, Industrial, Transportation, Residential/Commercial/Other, Solvents, Waste, International shipping, Aircraft | Coal, Oil, Gas, Cement. Bunker |

Table 3. Basic information of other emission datasets including GRACED.

As GRACED uses a top-down spatially gridding approach, errors and uncertainty in sectoral emission from Carbon Monitor, the national-level emission inventory, was evenly distributed in each grid. Considering additional uncertainties originating from spatial proxies of TROPOMI and EDGAR harnessed to distribute national-level emission, it is reasonable that the average grid-level uncertainty of GRACED is higher than certain national or global level emission inventories. Therefore, we also compared the uncertainty of GCP-GridFED, another gridded carbon emission dataset that published its uncertainty results, with the uncertainty of GRACED. The uncertainty value for GCP-GridFED was of 2018, as the latest uncertainty results of GCP-GridFED released by GCP team is of 2018. The GCP-GridFED showed a consistent uncertainty in the past six decades, ranging from 23.2% to 29.6%[27]. Therefore, using the uncertainty value of GCP-GridFED in 2018 does not impact the validity of this comparison. Comparing GRACED to GCP-GridFED, GRACED has a lower average uncertainty of ±21.0% (Fig. 6) during 2019-2021. As GCP-GridFED uncertainty is computed considering multiple variables[27], a similar calculation process with GRACED's, GRACED's lower uncertainty than GCP-GridFED indicates the higher accuracy of gridded carbon emission data that GRACED provides.

## Usage Notes

The GRACED2021 products[28] are now available at https://doi.org/10.6084/m9.figshare.21427437.v1 and also downloadable from our official website - https://carbonmonitor-graced.com. We have also provided an example of Python code to help users produce, read in and plot emissions for any grid in the dataset (https://github.com/jhong4776/GRACED2021). The emission data are stored in netCDF files per sector with the unit of kg carbon per hour (kgC/h) for each grid. For each sector (industry; power; residential consumption; ground transportation; international aviation; domestic aviation; international shipping), all daily files of the same month are merged into monthly *.zip files. The monthly *.zip file size is from 6 MB to 239 MB, respectively.

## Code Availability

Python code for producing, reading and plotting data in the dataset is provided at https://github.com/jhong4776/GRACED2021.

## Acknowledgements

Authors acknowledge Beijing Natural Science Foundation (JQ19032), the National Natural Science Foundation of China (grant 41921005 and grant 71874097), and the Qiu Shi Science & Technologies Foundation.

## Author contributions

XD and ZL designed the work and conducted the analysis. JH, YY, YH and YS conducted the analysis and contributed to editing the paper. PC, FC, YW, SJD, DH, XL, BZ, and ZD contributed to the design of methodology. FY and XS contributed to producing the figures. FC, MC, GJM, DG, ES, DC, PK, and HW contributed to the data collection. All authors contributed to discussing the scientific questions and editing the text.

## Competing interests

The authors declare no competing interests.